\documentclass[amsmath,aps,pra,reprint]{revtex4-1}

\usepackage[utf8]{inputenc}
\usepackage[T1]{fontenc}
\usepackage{graphicx} 
\usepackage{subfig}
\usepackage{hyperref} 
\usepackage{braket} 
\usepackage{siunitx} 
\usepackage{soul}
\usepackage{xcolor}

\newcommand{\orcid}[1]{\href{https://orcid.org/#1}{\includegraphics[width=10pt]{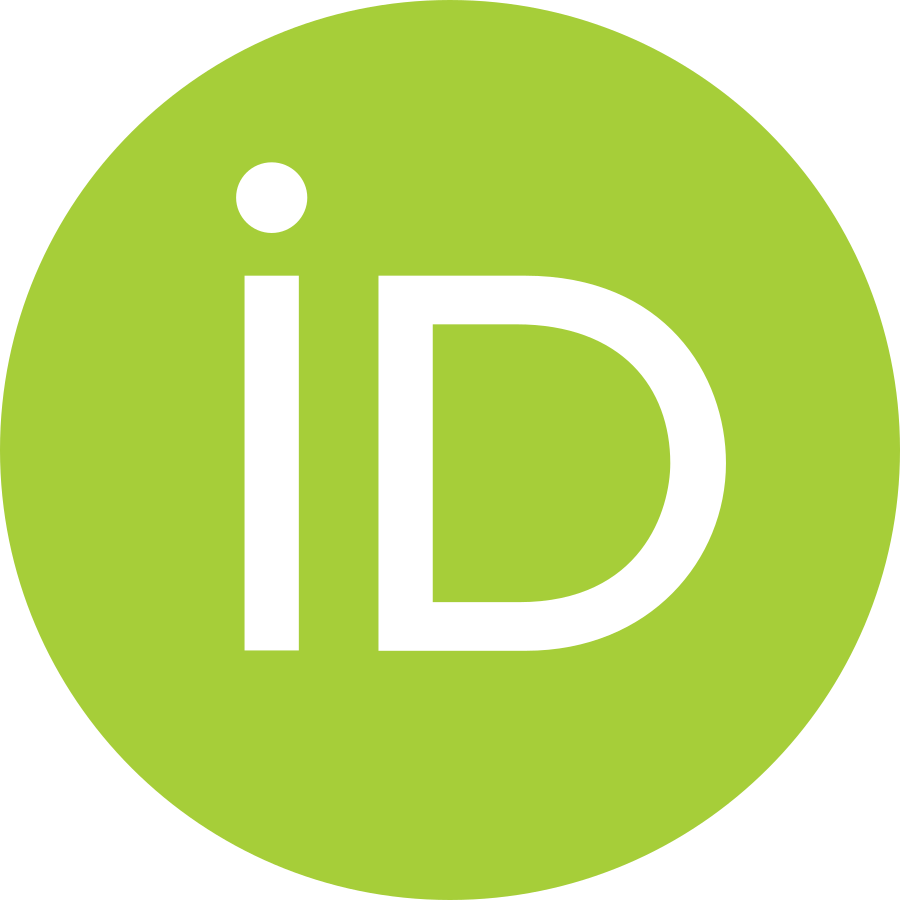}}}

\begin{document}

\title{Benchmarking of Fluorescence Lifetime Measurements \linebreak using Time-Frequency Correlated Photons}


\author{Tobias B. G\"{a}bler~\orcid{0000-0003-4874-428X}}
\email[corresponding author: ]{tobias.bernd.gaebler@iof.fraunhofer.de}
\affiliation{Fraunhofer Institute for Applied Optics and Precision Engineering IOF, Albert-Einstein-Straße 7, 07745 Jena, Germany}
\affiliation{Abbe Center of Photonics, Friedrich Schiller University Jena, Albert-Einstein-Straße 6, 07745 Jena, Germany}

\author{Patrick Then~\orcid{0009-0006-5103-7665}}
\affiliation{Institute of Applied Optics and Biophysics and Microverse Imaging Center, Friedrich Schiller University Jena, Philosophenweg 7,07743 Jena, Germany}
\affiliation{Leibniz Institute of Photonic Technology IPHT, Albert-Einstein-Straße 9, 07745 Jena, Germany}

\author{Christian Eggeling~\orcid{0000-0002-3698-5599}}
\affiliation{Institute of Applied Optics and Biophysics and Microverse Imaging Center, Friedrich Schiller University Jena, Philosophenweg 7,07743 Jena, Germany}
\affiliation{Leibniz Institute of Photonic Technology IPHT, Albert-Einstein-Straße 9, 07745 Jena, Germany}

\author{Markus Gr\"afe~\orcid{0000-0001-8361-892X}}
\affiliation{Institute of Applied Physics, Technical University of Darmstadt, Schloßgartenstraße 7, 64289 Darmstadt, Germany}

\author{Nitish Jain~\orcid{0009-0000-6055-3975}}
\affiliation{Fraunhofer Institute for Applied Optics and Precision Engineering IOF, Albert-Einstein-Straße 7, 07745 Jena, Germany}

\author{Valerio F. Gili~\orcid{0000-0001-8723-9986}}
\affiliation{Fraunhofer Institute for Applied Optics and Precision Engineering IOF, Albert-Einstein-Straße 7, 07745 Jena, Germany}

\date{\today}

\begin{abstract}
The investigation of fluorescence lifetime became an important tool in biology and medical science. So far, established methods of fluorescence lifetime measurements require the illumination of the investigated probes with pulsed or amplitude-modulated light. In this paper, we examine the limitations of an innovative method of fluorescence lifetime using the strong time-frequency correlation of entangled photons generated by a continuous-wave source. For this purpose, we investigate the lifetime of IR-140 to demonstrate the functional principle and its dependencies on different experimental parameters. We also compare this technique with state-of-the-art FLIM and observed an improved figure-of-merit. Finally, we discuss the potential of a quantum advantage.

\end{abstract}

\maketitle


\section{Introduction}
\label{sec:Introduction}
The capability of excitation and measurement of fluorescence light is fundamentally important for research activities in biology, chemistry and medicine. A variety of imaging methods base on fluorescence, starting from single or two-photon microscopy~\cite{Rumi2010} via fluorescence correlation spectroscopy~\cite{Krichevsky2002} towards super-resolution methods like stimulated emission depletion microscopy~\cite{Hell1994,Blom2017}. All these methods have benefits as well as drawbacks, which make them specialized for different applications.

One of the highly specialized methods is fluorescence lifetime imaging microscopy (FLIM). It represents a very valuable tool for investigations into the chemical environment of fluorescence dyes, especially if it is combined with a spectral analysis~\cite{Matayoshi1981,Baumann1985,Klitgaard2006,Becker2007,Knemeyer2007,Chorvat2009,Gerega2011,Tinnefeld2001}. However, a drawback of fluorescence lifetime measurements in comparison to standard fluorescence intensity detection is the requirement of pulsed light sources or optic modulators. This leads to high risk of photobleaching~\cite{Demchenko2020, Dasgupta2024} and additional technical effort. To avoid these drawbacks is an ongoing research topic.

In 2023, three research groups showed, independently of one another, that fluorescence lifetime measurements are possible using entangled photon pair sources~\cite{Harper2023,Eshun2023,Li2023}. Their work provides the basis for a new generation of fluorescence lifetime microscopes using upcoming quantum technologies. In particular, they utilized the time-frequency correlation of entangled photon pairs generated by spontaneous parametric down-conversion (SPDC). In doing so, one photon of this pair excites single-photon fluorescence, while the other one triggers the timing measurement. The time difference between the arrivals of heralding photon and the fluorescence photon represents a measure for the fluorescence lifetime. With this principle, they can avoid the need of pulsed lasers, but enhance the complexity due to the usage of an entangled photon pair source~\cite{Harper2023,Eshun2023}.

However, the mentioned studies~\cite{Harper2023,Eshun2023} were limited to the proof of concept and investigated setup-related influences only to a small extent. In this work, we consider different technical aspects of this method of lifetime measurements, such as the dependency on the spectrum of entangled photons or heralding efficiencies, to establish a framework for its implementation towards an application-oriented imaging method. To achieve this goal, we investigate the fluorescence dye IR-140 with photons generated by a SPDC source based on nonlinear waveguides. We further benchmark the results with respect to a classical state-of-the-art device. For this purpose, we will give insights into the methodology of fluorescence lifetime measurements using entangled photons (sec.~\ref{sec:Fundamentals}), describe our experimental approach (sec.~\ref{sec:Method}) and present our results (sec.~\ref{sec:Results}).

\section{Fundamentals}
\label{sec:Fundamentals}

While the measurement of fluorescence spectra provides insights into the electronic structures of matter, fluorescence lifetimes can provide information on the chemical environment and enable the separation of fluorophores with similar spectral emission~\cite{Berezin2010,Becker2012}.  For the determination of these lifetimes $\tau$, the decay of the fluorescence intensity $F$ over time $t$ after an excitation event will be considered.
\begin{eqnarray}
    F\left(t,\tau\right) = F_{0}\cdot\exp\left(-t/\tau\right)
    \label{eq:intensity decay}
\end{eqnarray}
$F_{0}$ represents the fluorescence intensity at $t=0$. Due to other processes that occur simultaneously, such as intersystem crossing or internal conversion, apart from fluorescence, observed lifetimes $\tau$ contain contributions of all these processes. For this reason, $\tau$ is composed by different decay constants $k_{i}$
\begin{eqnarray}
    \tau = \sum_{i} k_{i}^{-1}
    \label{eq:lifetime}
\end{eqnarray}
In general, eq.~\eqref{eq:lifetime} will be simplified by separation of the fluorescence process with constant $k_{f}$ and all non-radiative processes with constant $k_{nr}$. Both constants will be affected by the chemical environment, for example, by temperature, pH-value or the presence of other molecules specially in dense solutions~\cite{Chorvat2009}.

One of the state-of-the-art methods of determining $\tau$ is based on pulsed illumination~\cite{Berezin2010}. As the name suggests, pulsed light sources are used to excite the fluorescence. The time delay between an electronic trigger signal, which indicates the generation of a light pulse, and the fluorescence detection is a measure for the duration of the fluorescence process. Because of this direct consideration of time delays, FLIM based on pulsed illumination is referred to time-domain methods. However, the transmission of the trigger signal and the time of flight of excitation photons already introduce a time delay $\Delta t$. For this reason, the measured fluorescence signal $\widetilde{F}\left(t,\tau\right)$ represents a convolution of the real fluorescence decay $F$ and the so-called instrumental response function $IRF\left(t\right)$.
\begin{eqnarray}
    \widetilde{F}\left(t,\tau\right) = IRF\left(t\right) * F\left(t,\tau\right)
    \label{eq:convolution}
\end{eqnarray}
$IRF\left(t\right)$ describes, besides all time delays $\Delta t$ caused by the apparatus, especially the temporal uncertainty of the detection system which is determined by the jitter and, in case of pulsed illumination, the optical pulse width~\cite{Hirvonen2020}. $IRF$ will be measured by replacing the sample by a mirror or using fast decaying reference dyes~\cite{Luchowski2009}.

\begin{figure}
    \includegraphics[width=0.45\textwidth]{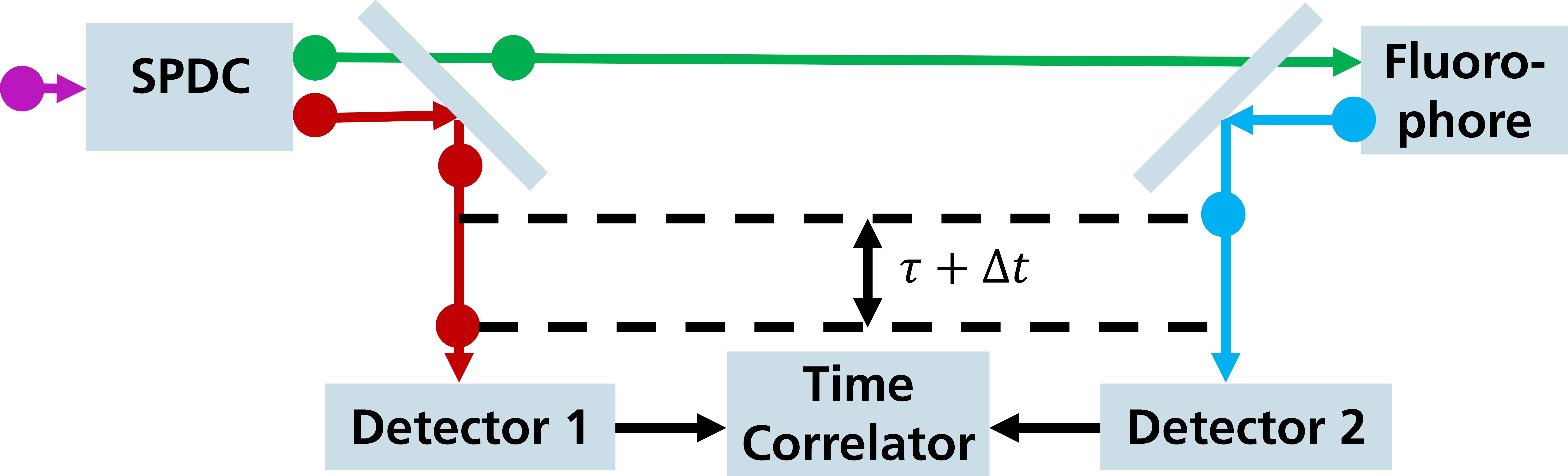}
    \caption{\label{fig:FLIM principle} Principle scheme of FLIM using time-frequency correlated photon pairs generated by SPDC}
\end{figure}

As shown simultaneously by Harper et al.~\cite{Harper2023} and Eshun et al.~\cite{Eshun2023}, the principle idea of time-domain FLIM can also be executed by CW light sources. Key ingredient is the usage of time-frequency correlated photon pairs, for example, generated by spontaneous parametric down-conversion (SPDC). As depicted in fig.~\ref{fig:FLIM principle}, the photons of a pair can be spatially separated in a deterministic way so that one photon excites the fluorescence dye whereas the other one triggers a time-correlated single photon counting unit (TCSPC). Because both photons of a pair are generated simultaneously (or more exactly within the Heisenberg uncertainty) during the SPDC process, the measured time difference between heralding and generated fluorescence photon corresponds to fluorescence lifetime $\tau$ and the additional time delay $\Delta t$ introduced by the apparatus.

Nevertheless, since this measurement principle corresponds to the  classical time-domain FLIM in essential features, the measured fluorescence decay $\widetilde{F}\left(t\right)$ can also be described via eq.~\eqref{eq:convolution}. Using TCSPC, $IRF$ and $\widetilde{F}\left(t\right)$ appear in form of temporal histograms. Because $IRF$ does not contain any temporal contributions, which lead to an asymmetric histogram, it can be described by a Gaussian distribution with mean value $\mu_{IRF}$ and standard deviation $\sigma_{IRF}$ given in eq.~\eqref{eq:IRF}~\cite{Chen2023}.
\begin{eqnarray}
    IRF\left(t\right) = \frac{1}{\sqrt{2\pi\sigma_{IRF}^2}}\cdot\exp{\left(-\frac{\left[t-\mu_{IRF}\right]^2}{2\sigma_{IRF}^2}\right)}
    \label{eq:IRF}
\end{eqnarray}
On the other side, $\widetilde{F}\left(t\right)$ is asymmetric because of the convolution of a symmetric with an asymmetric function. But the knowledge of eq.~\eqref{eq:intensity decay} and \eqref{eq:IRF} leads to an analytic expression of $\widetilde{F}\left(t\right)$~\cite{Chen2023}.
\begin{eqnarray}
    \widetilde{F}\left(t,\tau\right) &=& \frac{F_0}{2} \cdot \exp{\left(\frac{\sigma_{IRF}^2}{2\tau^2}\right)} \cdot \exp{\left(-\frac{t-\mu_{IRF}}{\tau}\right)} \nonumber\\
    & &\cdot \left[1+\text{erf}\left(\frac{t-\mu_{IRF}-\sigma_{IRF}^2/\tau}{\sqrt{2\sigma_{IRF}^2}}\right)\right]
    \label{eq:Ftilde}
\end{eqnarray}

These two expressions \eqref{eq:IRF} and \eqref{eq:Ftilde} finally enable the determination of $\tau$ from the measurement data as explained later in sec. \ref{sec:Data}.
Exploiting time-frequency correlated photon pairs for fluorescence life time measurements offers another advantage. The excitation wavelength is easily adjustable. A fact that was not yet considered. This adjustability directly stems from the momentum (phase-matching) and energy conservation conditions during SPDC processes. The wavelengths of the two photons of a generated pair are correlated and depend on the pump wavelength introduced into the nonlinear crystal as well as on its angle~\cite{Couteau2018} or temperature~\cite{Fedrizzi2007,Leon-Montiel2019}. Since, in particular, the crystal temperature can be easily modified, it may enable an use case for fluorescence lifetime spectroscopy. To implement this, photon pair sources with narrow SPDC bandwidth are necessary. Typical sources use nonlinear bulk crystals, which show relatively broad bandwidths depending on the length of the crystal~\cite{Fiorentino2007}. Better performance regarding narrow SPDC bandwidths is shown by photon pair sources based on nonlinear waveguides. Due to selective mode coupling, these sources have much narrower bandwidths~\cite{Fujii2007,Abolghasem2009} together with higher conversion efficiencies~\cite{Fiorentino2007,Spillane2007,Gaebler2024,Pollmann2024}.

Spectroscopic approaches to fluorescence lifetime measurements are usually known as sFLIM (spectrally resolved fluorescence lifetime imaging microscopy)~\cite{Becker2007,Chorvat2009,Knemeyer2007,Tinnefeld2001}. This special type of FLIM usually investigates the dependency of lifetime $\tau$ on the fluorescence emission wavelength. However, the analysis regarding the excitation wavelength is also practiced in special cases, for example, if multiple spectrally separated fluorophores are present in the probe~\cite{Volz2018}. The physical basis for sFLIM is the wavelength dependency of decay constants $k_i$ caused by diverse effects. The most obvious one are the electronic structures of dye molecules which result in characteristic wavelength-dependent absorption and fluorescence behavior of every fluorescence species. However, these electronic structures can be influenced by, for example, chemical bonds to the chemical environment.

\section{Method}
\label{sec:Method}

\subsection{Experimental Setup}
\label{sec:setup}

\begin{figure*}[tpb]
    \includegraphics[width=0.9\textwidth]{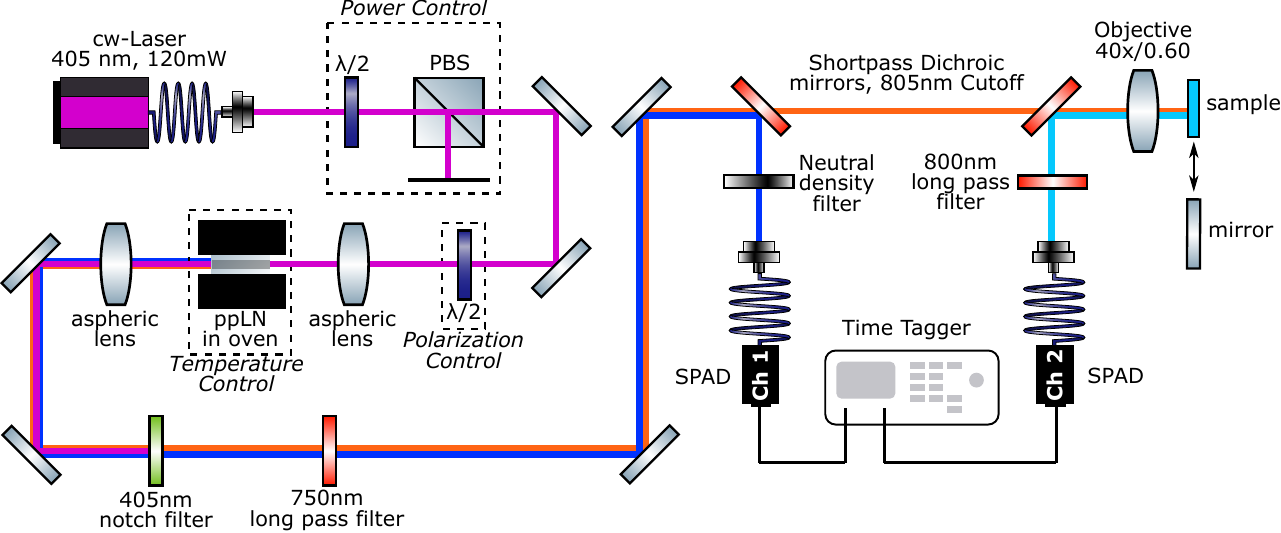}
    \caption{\label{fig:setup} Experimental Setup for the measurement of fluorescence lifetimes with time-frequency correlated photons. The left part shows the photon pair source, whereas the right side is the tailor-built microscope. Violet indicates the optical path of the pump, dark blue the heralding, orange the excitation and light blue the fluorescence beam.}
\end{figure*}

Our experimental setup consisted of a source of time-frequency correlated photon pairs and a tailor-built microscope. The photon pairs source, shown on the left side of fig.~\ref{fig:setup}, based on a periodically poled lithium-niobate waveguide with a length of $\SI{20}{\milli\metre}$ (AdVR), which was pumped by a CW diode laser with a center wavelength of $\SI{405}{\nano\metre}$  (Toptica iBeam-Smart-405-S-HP). Correlated photons were generated by SPDC type-0 with a center wavelength of $\SI{810}{\nano\metre}$. Residual pump photons were filtered out by $\SI{405}{\nano\metre}$-notch (Thorlabs NF405-13) and $\SI{750}{\nano\metre}$-longpass filters (Thorlabs FELH0750). A detailed description of the key characteristics of this source is given in ref.~\cite{Gaebler2024}.

The tailor-built microscope, shown on the right side of fig.~\ref{fig:setup}, contained two shortpass dichroic mirrors (Thorlabs DMSP805R) and a microscope objective (Olympus LUCPLFLN40X). The first dichroic mirror enabled the separation of heralding photons (above $\SI{805}{\nano\metre}$) and exciting photons (below $\SI{805}{\nano\metre}$). Heralding photons were coupled by a adjustable collimator (Thorlabs PAF2P-A15B) into a multimode fiber (Thorlabs M123L01). This fiber is connected to a single-photon avalanche detecter (Excelitas SPCM-800-42-FC with a timing jitter of $\SI{350}{\pico\second}$ at a wavelength of $\SI{825}{\nano\metre}$; channel 1). For avoiding the saturation of this detector, the pump power coupled into the waveguide was set to $\SI[separate-uncertainty = true]{27.6(1.3)}{\micro\watt}$ to limit the total number of generated photon pairs. Additional neutral density filters (Thorlabs NUK01) were added in front of the collection of heralding photons to investigate the dependence on heralding efficiency $\eta$.

Photons with a wavelength below $\SI{805}{\nano\metre}$, which passed the first dichronic mirror, were led through the second mirror into the microscope objective and focused on the sample to excite the fluorescence dye. Fluorescence photons emitted with a wavelength above $\SI{805}{\nano\metre}$ are collected by this objective and redirected by the second dichroic mirror into a fiber coupler (Thorlabs PAF2-A7B). Due to the imperfect reflectivity of the dichroric mirrors, an additional $\SI{800}{\nano\metre}$-longpass filter (Thorlabs FELH0800) was used to prevent that residual heralding or exciting photons are reaching this light path. Fluorescence photons are guided with a multimode fiber (Thorlabs M123L01) to a second detector (also Excelitas SPCM-800-42-FC; channel 2). Both detectors, channel 1 and 2, are electrically connected to a TCSPC (QuTools QuTag Standard with timing jitter of $\SI{3.0}{\pico\second}$ (RMS) and a resolution of $\SI{1.0}{\pico\second}$).

\begin{figure}[tpb]
    \includegraphics[width=0.45\textwidth]{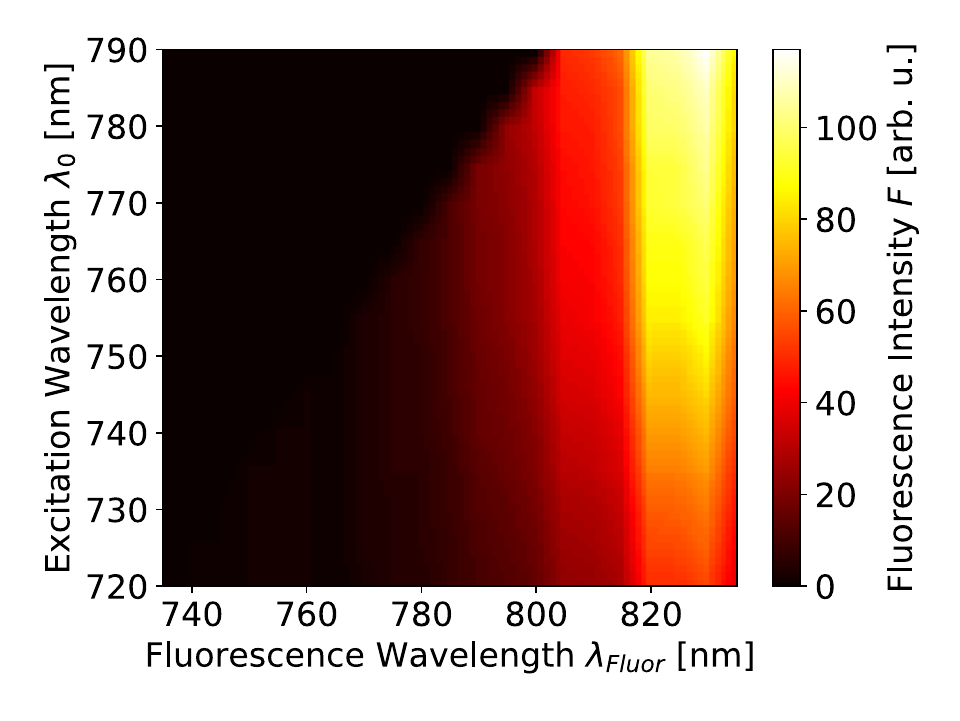}
    \caption{\label{fig:LambdaLambdaScan} Excitation vs fluorescence emission wavelength for the used sample of IR-140 measured by a Leica Stellaris 8.}
\end{figure}

As fluorescence sample, the fluorophore IR-140 (Sigma-Aldrich 260932-100MG) with a reported quantum yield of $0.167$~\cite{Rurack2011} was used in a solution with ethanol (Sigma-Aldrich 51976-500-ML-F) with a concentration of $\SI{1.3}{\milli\mol\per\litre}$. This sample shows strong fluorescence emission above $\SI{820}{\nano\metre}$ for excitation light below $\SI{790}{\nano\metre}$ (see fig.~\ref{fig:LambdaLambdaScan}). Thus, it meets the aforementioned conditions. For the following lifetime measurements, the dye was filled into a cuvette (Thorlabs CV1Q035AE) and mounted on a two-axis translation stage (stack of two Thorlabs PT1). On the same stage, a mirror (Thorlabs BB1-E03) was mounted to allow a simple exchange with the sample in the optical path to record the $IRF$ following the methodology of ref.~\cite{Harper2023}. 

It must be mentioned that the usage of a mirror can distort the $IRF$. Because of the different excitation and fluorescence wavelengths, the photon detection efficiency as well as the jitter varies between the measurements of $IRF$ and $\widetilde{F}$ and, thus, impair their compatibility. This can be avoided by using ultra-fast decaying dyes as reference in the measurement of $IRF$~\cite{Luchowski2009}. However, we assume that these errors play a subordinate role in our case. The photon detection efficiencies between excitation and fluorescence wavelengths differ by approximately $\SI{10}{\percent}$, which will influence the amplitudes of $IRF$ and $\widetilde{F}$. In contrast, the determination of $\tau$ does not directly depend on these amplitudes as visible in eq.~\eqref{eq:IRF} and~\eqref{eq:Ftilde}. Moreover, the documented timing jitter is given for $\SI{825}{\nano\metre}$, which is nearly in the center between the excitation and fluorescence wavelength of IR-140. For this reason, we also assume that the jitter does not vary significantly between $IRF$ and $\widetilde{F}$.

\begin{figure}[tpb]
    \includegraphics[width=0.45\textwidth]{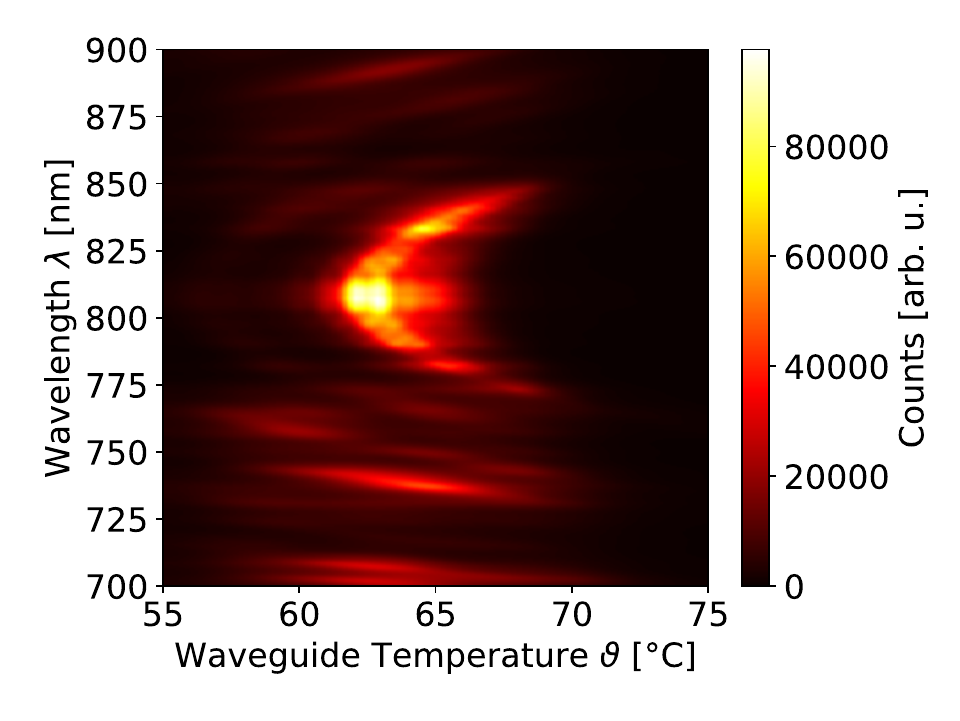}
    \caption{\label{fig:Temperature-vs-Wavelength} Spectrum of the photon pair source depicted in fig.~\ref{fig:setup} for different waveguide temperatures $\vartheta$ and measured by a Ocean Insight QE Pro.}
\end{figure}

As explained in sec.~\ref{sec:Fundamentals}, the temperature dependent recording of the histograms enables a direct correlation to the excitation wavelength $\lambda_0$. This relation between waveguide temperature $\vartheta$ and wavelength of time-frequency correlated photons is shown in fig.~\ref{fig:Temperature-vs-Wavelength}. Due to the conservation of energy and momentum, the described source exhibits a degenerate spectrum at approximately $\vartheta=\SI{62}{\celsius}$. Above this temperature, the spectra have two significant peaks with correlated center wavelengths, representing both photons of a pair. For this reason, waveguide temperatures above $\vartheta=\SI{63}{\celsius}$ were used exclusively to ensure a clear separation of heralding and exciting photons in all measurements.

\subsection{Data Recording and Processing}
\label{sec:Data}

The TCSPC was set to a bin width of $\SI{2}{\pico\second}$ and bin count of $N_t=5000$ to ensure the complete recording of all histograms with high temporal precision. All histograms of $IRF$ and $\widetilde{F}$, as well as the single and coincidence counts, are measured for different cases: First, the integration time $T$ was increased from $\SI{1}{\second}$ to $\SI{2}{\second}$, $\SI{5}{\second}$, $\SI{10}{\second}$, $\SI{20}{\second}$, $\SI{30}{\second}$, $\SI{1}{\minute}$, $\SI{2}{\minute}$, $\SI{4}{\minute}$, $\SI{5}{\minute}$, $\SI{10}{\minute}$, $\SI{15}{\minute}$, $\SI{30}{\minute}$ and $\SI{1}{\hour}$ for a fixed waveguide temperature of $\vartheta=\SI{64}{\celsius}$. This case is used to investigate the accuracy of the lifetime determination as it was already shown in ref.~\cite{Harper2023} (section~\ref{sec:IntTime}). Secondly, measurements with different neutral density filters (optical densities of $0.3$, $0.6$, $1.0$, $1.5$, $2.0$ and $4.0$) placed in front of the heralding detector (channel 1) for an integration time of $T=\SI{15}{\minute}$ and a waveguide temperature of $\vartheta=\SI{64}{\celsius}$ were executed. The aim of this is the consideration of the effect of heralding efficiency and signal-to-noise ratio on the lifetime determination (section~\ref{sec:HeraldingEfficiency}). The last measurement was performed to investigate the possibility of spectroscopic applications (section~\ref{sec:Temperature}). For this purpose, the waveguide temperature $\vartheta$ was varied between $\SI{63.0}{\celsius}$ and $\SI{70.0}{\celsius}$ for a fixed integration time of $\SI{15}{\minute}$. 

To reduce the influence of random noise and accidental changes during the execution of the experiments, every single histogram was measured multiple times under the same setting. In particular, the measurements at different integration times were performed three times, for different neutral density filters two times and for different waveguide temperatures four times. During data analysis, these multiple datasets were used to calculate the average of coincidence events for every time bin in the respective histogram (step~\ref{item:averaging} in the following list).

Because of the long measurement times, TCSPC and temperature controller were controlled automatically by Python codes. These are available in ref.~\cite{Gaebler2025}. The post-processing of the histograms was also performed by Python codes in the following way:
\begin{enumerate}
  \item Elimination of the background of $IRF\left(t\right)$ and $\widetilde{F}\left(t\right)$ caused by accidental coincidence events, for example, resulting from erroneous correlation of two random background photons.
  \item\label{item:averaging} Time bin wise averaging of $IRF\left(t\right)$ and $\widetilde{F}\left(t\right)$ over the sets of measurement
  \item Shift $IRF\left(t\right)$ so that $\mu_{IRF}=\SI{0}{\pico\second}$ for all measurements.
  \item Shift $\widetilde{F}\left(t\right)$ equally to their corresponding $IRF\left(t\right)$
  \item Normalize $IRF\left(t\right)$ and find value of $\sigma_{IRF}$ by curve-fitting using eq.~\eqref{eq:IRF}
  \item Find $\tau$ by curve-fitting of $\widetilde{F}\left(t\right)$ using eq.~\eqref{eq:Ftilde}
\end{enumerate}

\section{Results}
\label{sec:Results}

\subsection{Integration Time}
\label{sec:IntTime}

\begin{figure*}[htpb]
    \centering
    \subfloat[]{\includegraphics[width=0.40\textwidth]{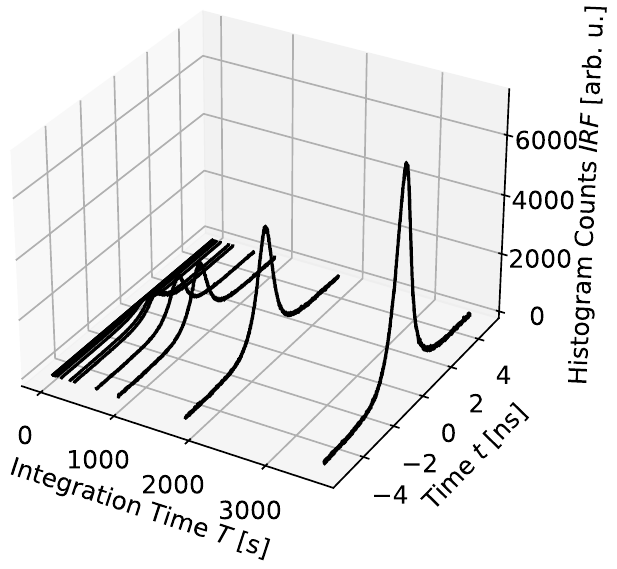}\label{fig:IRF IntTime}}
    \hspace{1cm}
    \subfloat[]{\includegraphics[width=0.40\textwidth]{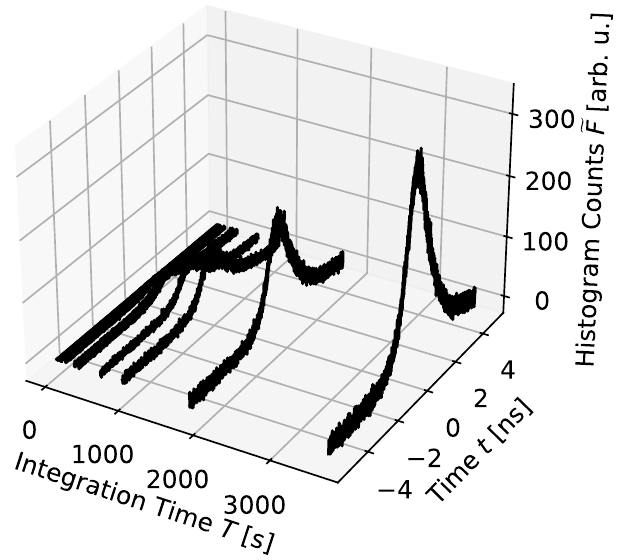}\label{fig:F IntTime}}
    \caption{$IRF$ (\ref{fig:IRF IntTime}) and $\widetilde{F}$ (\ref{fig:F IntTime}) for different integration times $T$}
    \label{fig:histograms IntTime}
\end{figure*}

\begin{figure*}[htpb]
    \centering
    \subfloat[]{\includegraphics[width=0.45\textwidth]{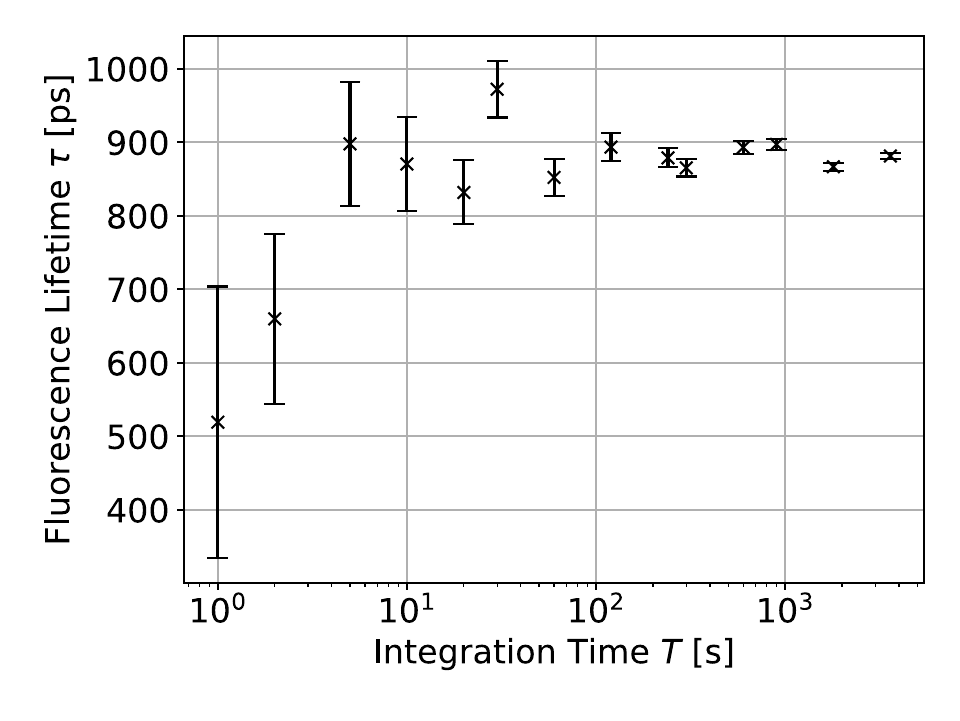}\label{fig:tau IntTime}}
    \hspace{1cm}
    \subfloat[]{\includegraphics[width=0.45\textwidth]{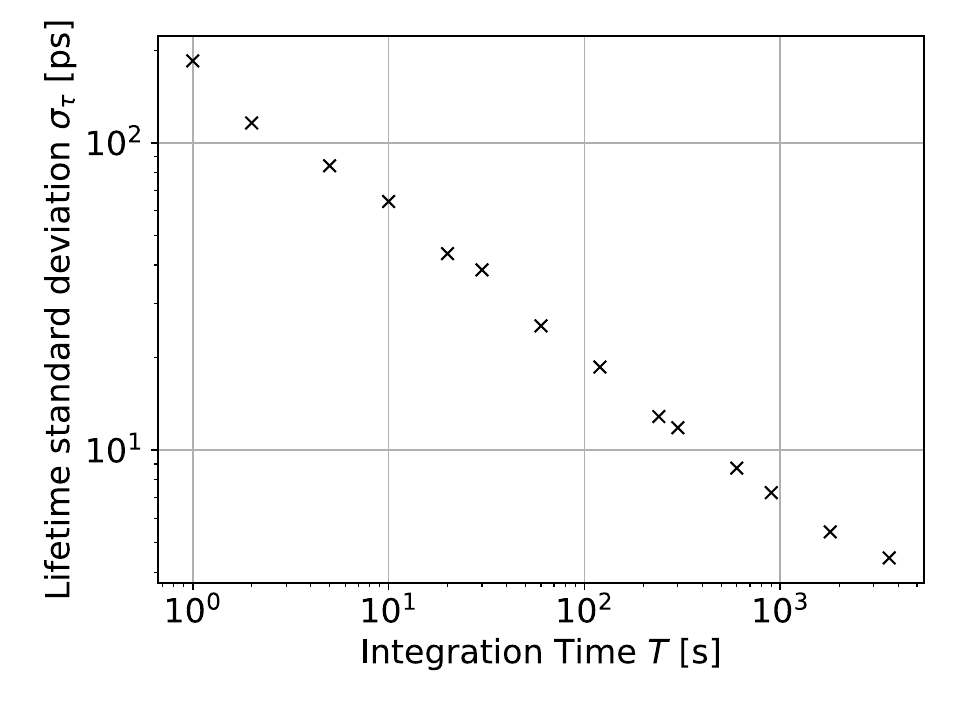}\label{fig:std tau IntTime}}
    \caption{Fluorescence lifetime $\tau$ (\ref{fig:tau IntTime}) and its standard deviation $\sigma_\tau$ (\ref{fig:std tau IntTime}) depending on the integration time $T$}
    \label{fig:Lifetime IntTime}
\end{figure*}

The averaged and background-corrected histograms for different integration times $T$ are shown in fig.~\ref{fig:histograms IntTime}. As expected, the histogram peaks become more significant with longer integration times $T$. Furthermore, $\widetilde{F}$ is more noisy than $IRF$ caused by the much lower coincidence count rate. However, the determined fluorescence lifetime of IR-140 is stabilized for longer integration times $T$ at around $\SI{885}{\pico\second}$ (fig.~\ref{fig:tau IntTime}) and its standard deviation drops into the range of $\SI{3}{\femto\second}$ (fig.~\ref{fig:std tau IntTime}).

\begin{figure}[tpb]
    \includegraphics[width=0.45\textwidth]{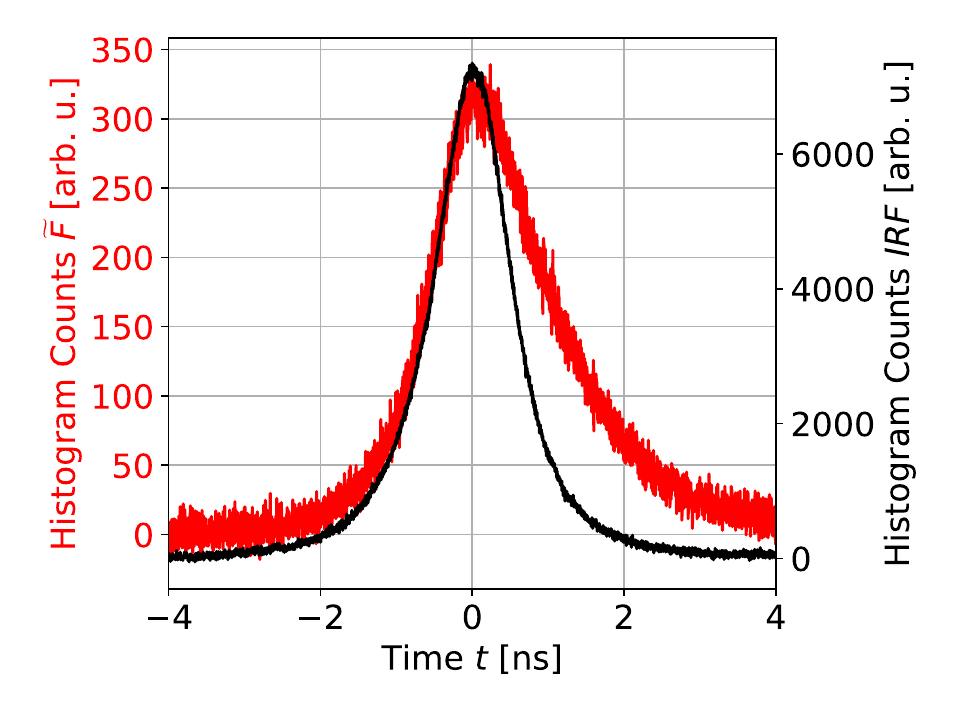}
    \caption{\label{fig:Comparison-Histogram} Histograms $IRF\left(t\right)$ and $\widetilde{F}\left(t,\tau\right)$ for the integration time $T=\SI{1}{\hour}$. The asymmetric broadening indicates the fluorescence decay.}
\end{figure}

Nevertheless, it is remarkable that the values of $\tau$ diverge strongly for short integration times $T$. It is likely that this effect depends on the significance of the histogram and, thus, amount of detected coincidences. To gain a deeper understanding of this effect, the influence of the signal-to-noise ratio and the heralding efficiencies is investigated by inserting neutral density filters into the optical path of channel 1. The gained results are shown in the following subsection.

\subsection{SNR and Heralding Efficiencies}
\label{sec:HeraldingEfficiency}

Due to the background correction of all histograms, we use the following definition of the signal-to-noise ratio $SNR$~\cite{Gili2022}.
\begin{eqnarray}
    SNR_{IRF} &=& \frac{IRF^{\textbf{max}}}{\sigma_{IRF}}\\
    SNR_{\widetilde{F}} &=& \frac{\widetilde{F}^{\textbf{max}}}{\sigma_{\widetilde{F}}}
    \label{eq:SNR}
\end{eqnarray}

$IRF^{max}$ and $\widetilde{F}^{max}$ represent the peak values of the non-normalized histograms. $\sigma_{IRF}$ and $\sigma_{\widetilde{F}}$ are the standard deviation of their background noise. Their estimation is based on the first $N_t=300$ bins at times $t_i$ of every recorded histogram, which corresponds to a time range of $\SI{600}{\pico\second}$. It ensures that all involved bins are related to the background noise and not to the relevant histogram range, which contains information about $\Delta t$ or $\tau$.

\begin{eqnarray}
    \sigma_{IRF} &=& \sqrt{\frac{1}{N_t}\cdot\sum_{i=0}^{N_t}IRF^2\left(t_i\right)}\\
    \sigma_{\widetilde{F}} &=& \sqrt{\frac{1}{N_t}\cdot\sum_{i=0}^{N_t}\widetilde{F}^2\left(t_i\right)}
    \label{eq:RMS}
\end{eqnarray}

\begin{figure*}[htpb]
    \centering
    \subfloat[]{\includegraphics[width=0.45\textwidth]{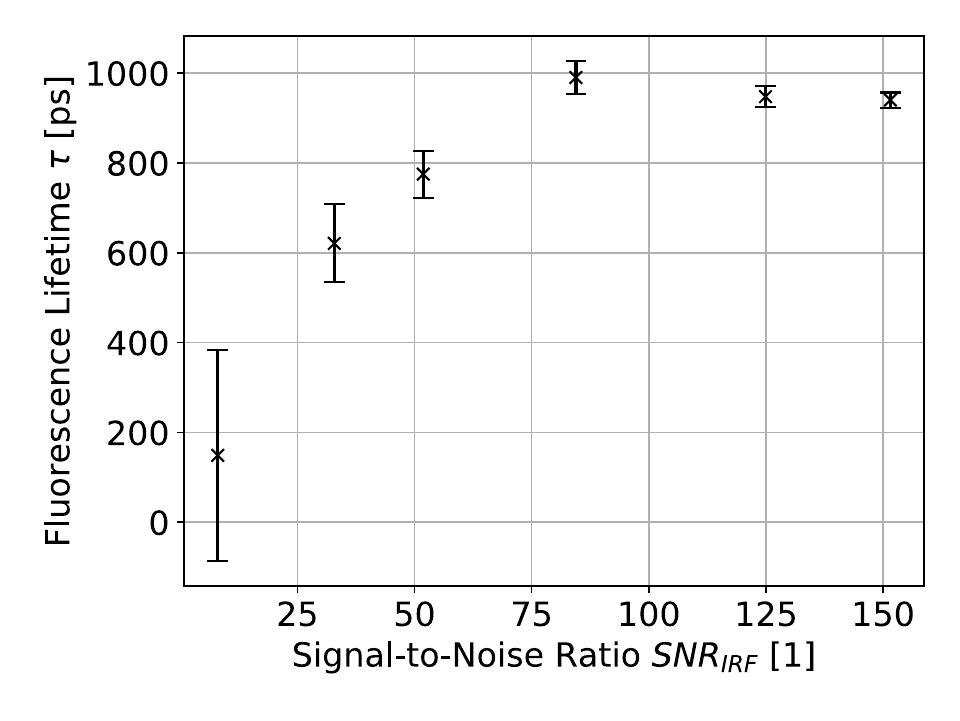}\label{fig:SNR IRF}}
    \hspace{1cm}
    \subfloat[]{\includegraphics[width=0.45\textwidth]{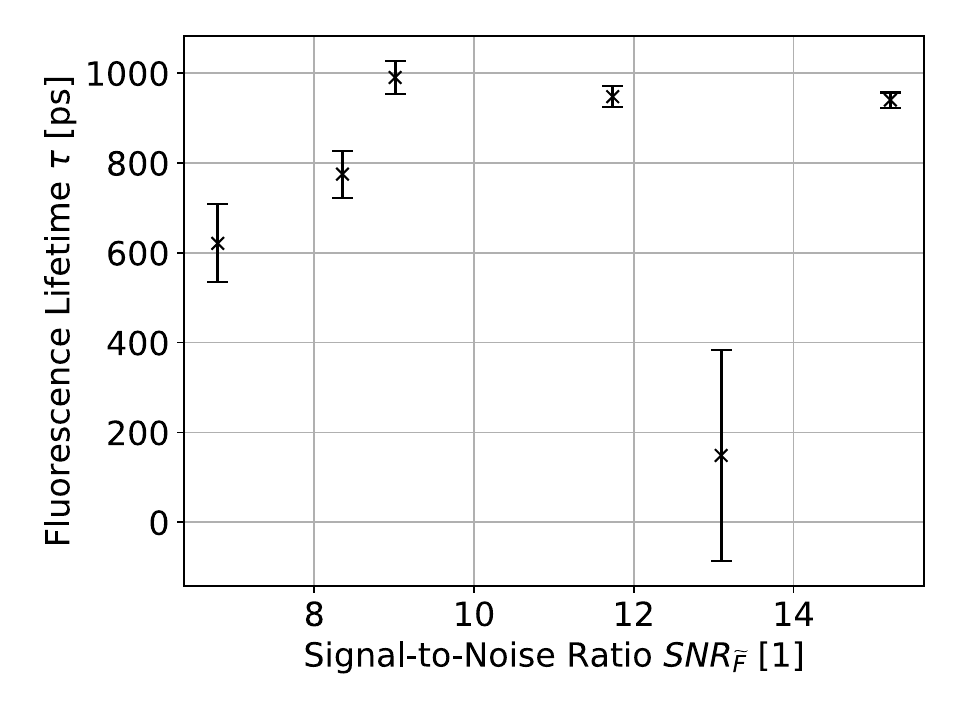}\label{fig:SNR F}}
    \caption{Effect of the signal-to-noise ratios $SNR_{IRF}$ (\ref{fig:SNR IRF}) and $SNR_{\widetilde{F}}$ (\ref{fig:SNR F}) on fluorescence lifetime $\tau$}
    \label{fig:SNR}
\end{figure*}

Fig.~\ref{fig:SNR} shows the fluorescence lifetimes over the signal-to-noise ratio of instrument response function $SNR_{IRF}$ (fig.~\ref{fig:SNR IRF}) and measured fluorescence $SNR_{\widetilde{F}}$ (fig.~\ref{fig:SNR F}). In case of the instrument response function, $\tau$ becomes stable for $SNR_{IRF}\gtrsim80$, whereas it is $SNR_{\widetilde{F}}\gtrsim9$ in case of the fluorescence measurement. Only the data point at $SNR_{\widetilde{F}}\approx13.1$, which corresponds to the measurement with a neutral density filter with optical density of $4.0$, differs strongly. The reason for this is that no clear peak in the histogram was visible anymore and, thus, $\widetilde{F}^{\textbf{max}}$ represents the largest noise count in this case.

To conclude, fig.~\ref{fig:SNR} shows that we can access a regime of reliable measurements, but with the clarity that the signal-to-noise ratio $SNR_{\widetilde{F}}$ of the fluorescence detection is the most relevant criterion. This is evident since the amount of fluorescence photons is much lower than the amount of photons collected by the heralding detector or by the fluorescence detector during the measurement of the instrument response function.

The signal-to-noise ratio, which is a standard performance indicator in data processing, only gives information about required coincidence counts vs. accidental coincidences in the recorded histograms for reliable measurements. Since FLIM with correlated photon pairs is based on coincidence detection of two light beams, heralding efficiencies $\eta$ may offer additional lower bounds on detection efficiencies and losses in light of reliable measurements. The heralding efficiencies $\eta_{1,2}$ are defined as ratio of measured coincidence rate $R^{\textbf{coin}}$ to single count rates $R^{\textbf{single}}_{1,2}$ at detection channel 1 or 2.
\begin{eqnarray}
    \eta_{1,2} &=& \frac{R^{\textbf{coin}}}{R^{\textbf{single}}_{1,2}}
    \label{eq:heraldingEfficiency}
\end{eqnarray}

\begin{figure*}[htpb]
    \centering
    \subfloat[]{\includegraphics[width=0.45\textwidth]{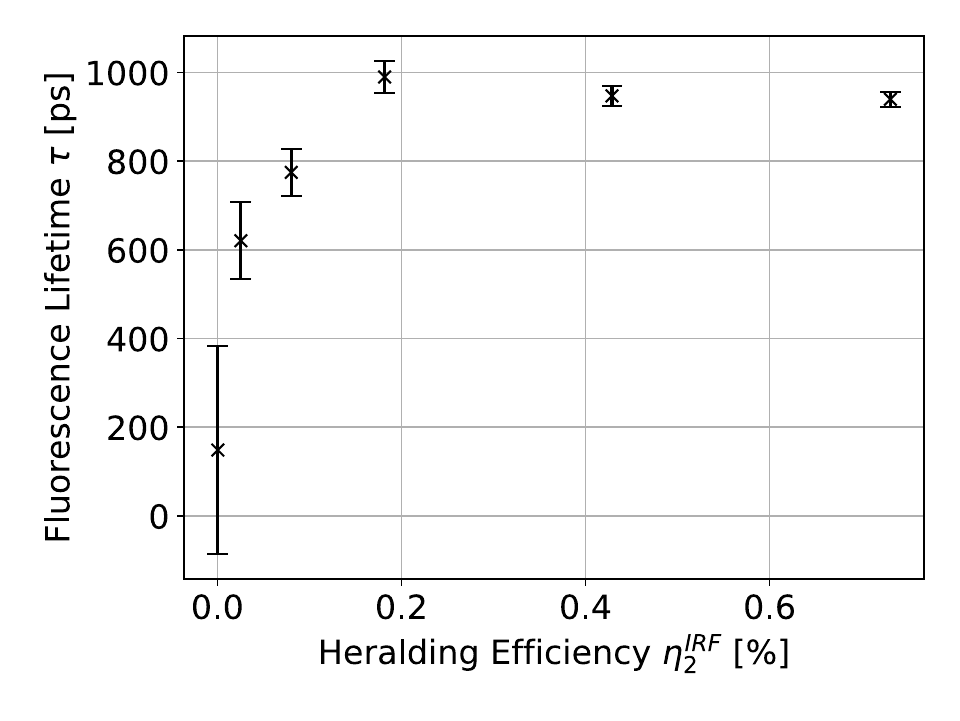}\label{fig:eta_2 IRF}}
    \hspace{1cm}
    \subfloat[]{\includegraphics[width=0.45\textwidth]{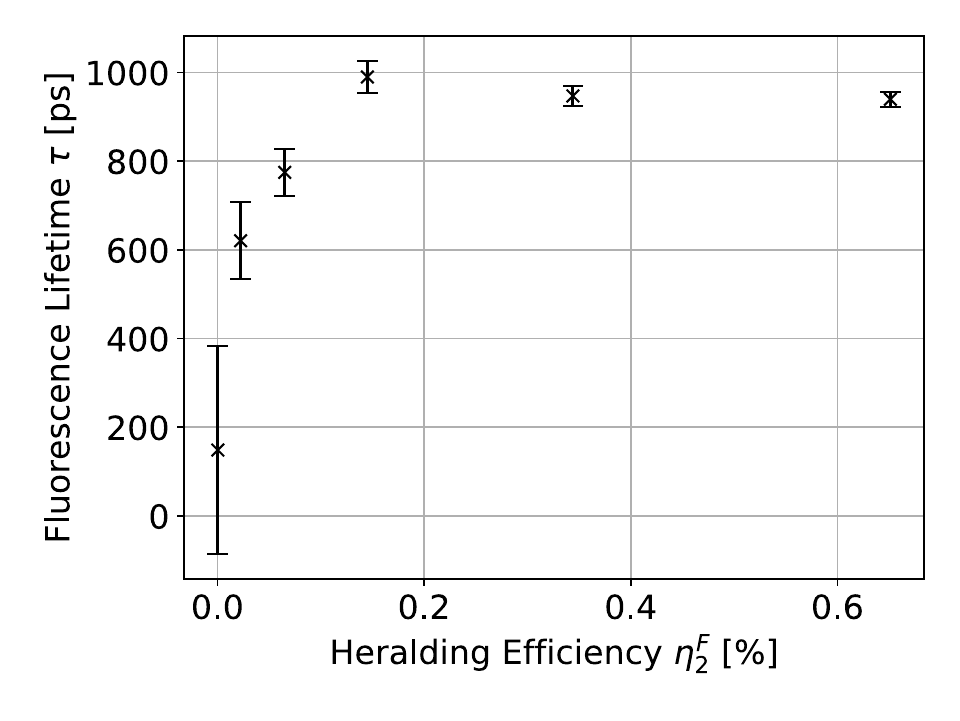}\label{fig:eta_2 F}}
    \caption{Effect of the heralding efficiencies $\eta_2^{IRF}$ (\ref{fig:eta_2 IRF}) and $\eta_2^{F}$ (\ref{fig:eta_2 F}) regarding channel 2 (fluorescence detector) on fluorescence lifetime $\tau$}
    \label{fig:eta_2}
\end{figure*}

Fig.~\ref{fig:eta_2} illustrates the fluorescence lifetimes $\tau$ depending on the heralding efficiencies $\eta_2$ with regard to the photon counts $R^{\textbf{single}}_{2}$ at the fluorescence detector (channel 2) for $IRF$ and $\widetilde{F}$. As visible, the values become stable for heralding efficiencies $\eta_2\gtrsim\SI{0.1}{\percent}$. In comparison, entangled photon pair sources usually show heralding efficiencies around $\SI{40}{\percent}$~\cite{Ramelow2013,Steinlechner2014,Brambila2023}. As a consequence, time-domain FLIM with time-frequency correlated photon pairs is feasible even if the ideal heralding efficiency of a photon pair source cannot be reached because of the lossy conversion of one beam to fluorescence photons.

\begin{figure*}[htpb]
    \centering
    \subfloat[]{\includegraphics[width=0.45\textwidth]{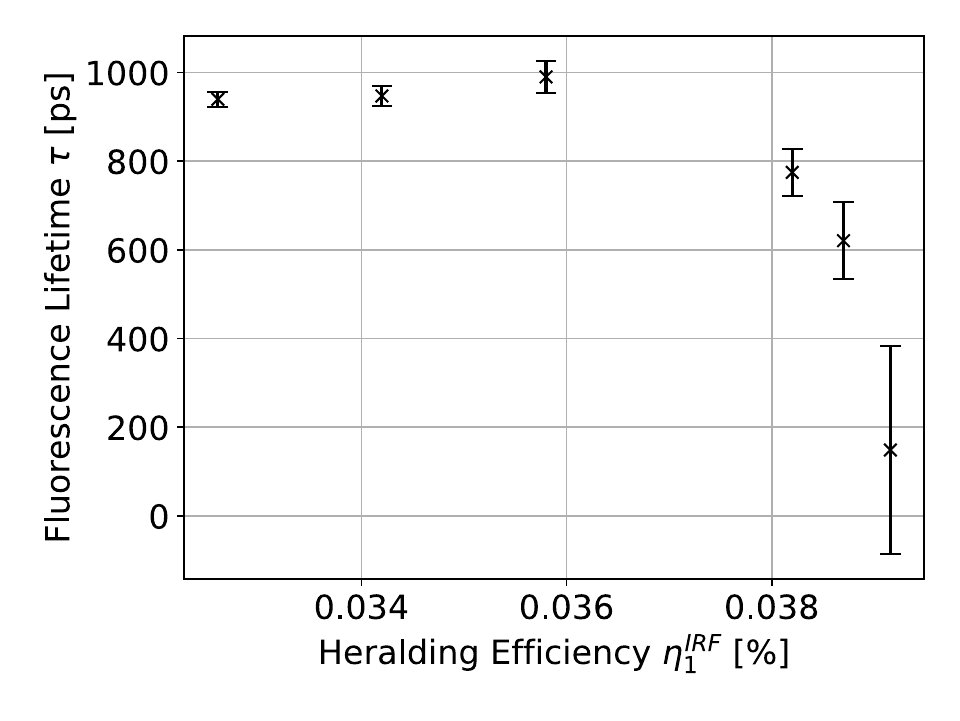}\label{fig:eta_1 IRF}}
    \hspace{1cm}
    \subfloat[]{\includegraphics[width=0.45\textwidth]{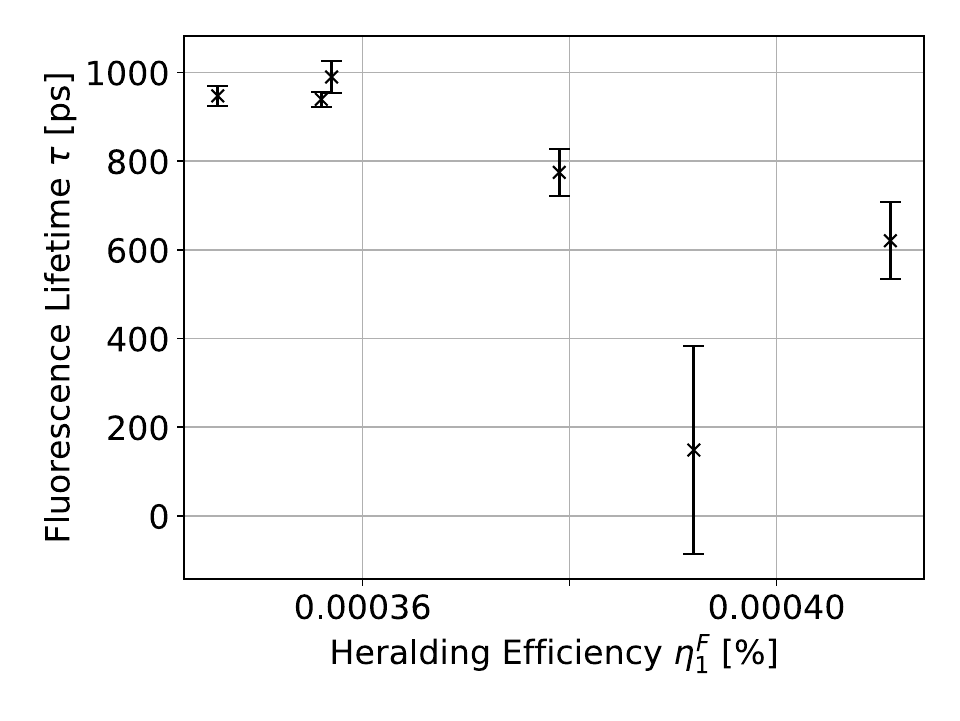}\label{fig:eta_1 F}}
    \caption{Effect of the heralding efficiencies $\eta_1^{IRF}$ (\ref{fig:eta_1 IRF}) and $\eta_1^{F}$ (\ref{fig:eta_1 F}) regarding channel 1 (heralding detector) on fluorescence lifetime $\tau$}
    \label{fig:eta_1}
\end{figure*}

On the other side, the heralding efficiency $\eta_1$ with regard to heralding detection (channel 1) does not play an important role. As depicted in fig.~\ref{fig:eta_1}, the fluorescence lifetime $\tau$ decreases with $\eta_1$ at certain points. Since in general the single photon rate $R^{\textbf{single}}_{1}$ is much higher than $R^{\textbf{single}}_{2}$, its reduction, for example, by neutral density filters, has only little influence on the coincidence rate $R^{\textbf{coin}}$. For this reason, the heralding efficiency $\eta_1$ increases with higher optical densities, and $\tau$ shows the opposite behavior in comparison to fig.~\ref{fig:eta_2}.

This technique shows promising potential for future application-oriented developments, including its suitability for advanced quantum-enhanced biological imaging. The current acquisition time of $\SI{15}{\minute}$ enables precise measurements; however, to make the approach more feasible for biomedical imaging applications requiring fast frame rates, enhancements in performance would be beneficial. Specifically, increasing the count rate of detected fluorescence in channel 2 relative to the heralding photons in channel 1 could be achieved through further optimization of the optical system, for example, by oil immersion objectives with higher numerical aperture or free-space detectors. While increasing the SPDC generation rate via higher pump power might initially seem like a viable option, this approach requires careful consideration, as the linear increase in photon-pair rates can lead to detector saturation~\cite{Gaebler2024}, resulting in artifacts and inaccuracies in fluorescence lifetime measurements. These insights provide a roadmap for advancing the method and enhancing its applicability in fast-paced imaging scenarios.

\subsection{Spectroscopic Approach}
\label{sec:Temperature}

Because of the usage of a waveguide source with a narrow bandwidth in comparison to usual photon pair sources based on nonlinear bulk crystals, a spectroscopic use case becomes conceivable. For this purpose, the waveguide temperature was varied to change the excitation wavelength according to fig.~\ref{fig:Temperature-vs-Wavelength}. 

\begin{figure*}[htpb]
    \centering
    \subfloat[]{\includegraphics[width=0.40\textwidth]{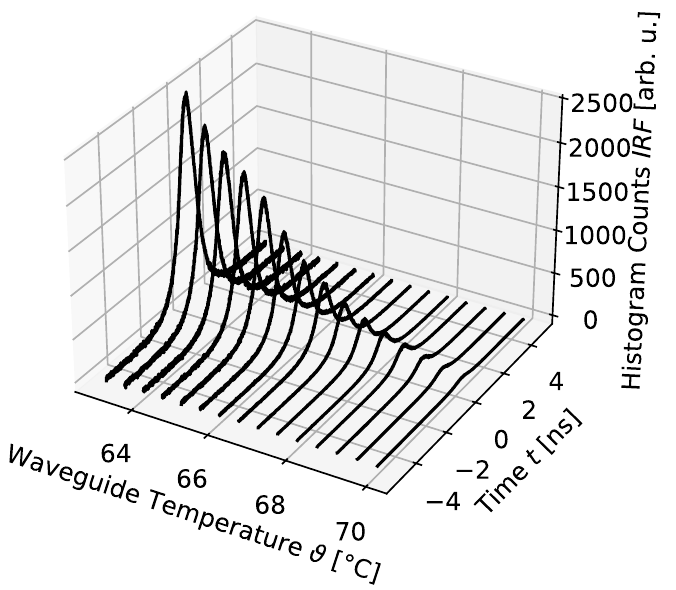}\label{fig:IRF Temperature}}
    \hspace{1cm}
    \subfloat[]{\includegraphics[width=0.40\textwidth]{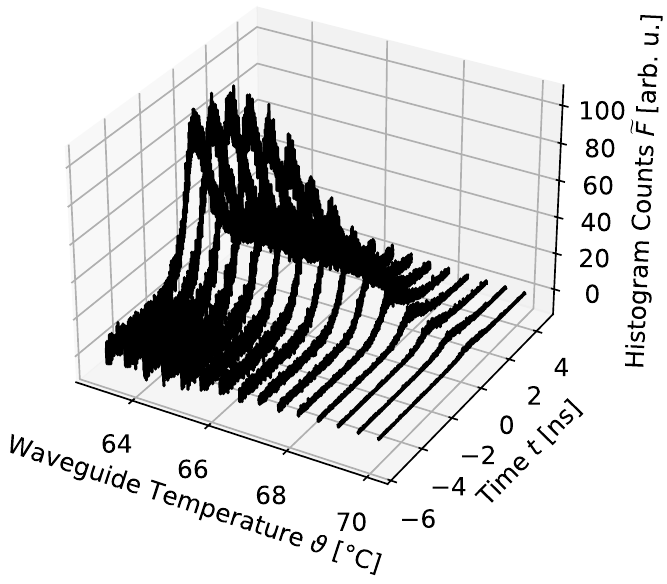}\label{fig:F Temperature}}
    \caption{$IRF$ (\ref{fig:IRF Temperature}) and $\widetilde{F}$ (\ref{fig:F Temperature}) for different waveguide temperatures $\vartheta$}
    \label{fig:histograms Temperature}
\end{figure*}

Fig.~\ref{fig:histograms Temperature} shows the histograms $IRF$ and $\widetilde{F}$ for different waveguide temperatures $\vartheta$. In general, the counts are falling for increased temperature, which corresponds to the decrease in the number of photons generated by the SPDC process for higher temperatures (see fig.~\ref{fig:Temperature-vs-Wavelength}). However, $\widetilde{F}$ shows a deviation from this general behavior: The highest amount of coincidence counts appear at $\vartheta=\SI{64}{\celsius}$. This indicates an optimum waveguide temperature $\vartheta$ where the combination of excitation wavelength and SPDC generation rate achieves the highest possible fluorescence rate.

\begin{figure}[tpb]
    \centering
    \includegraphics[width=0.45\textwidth]{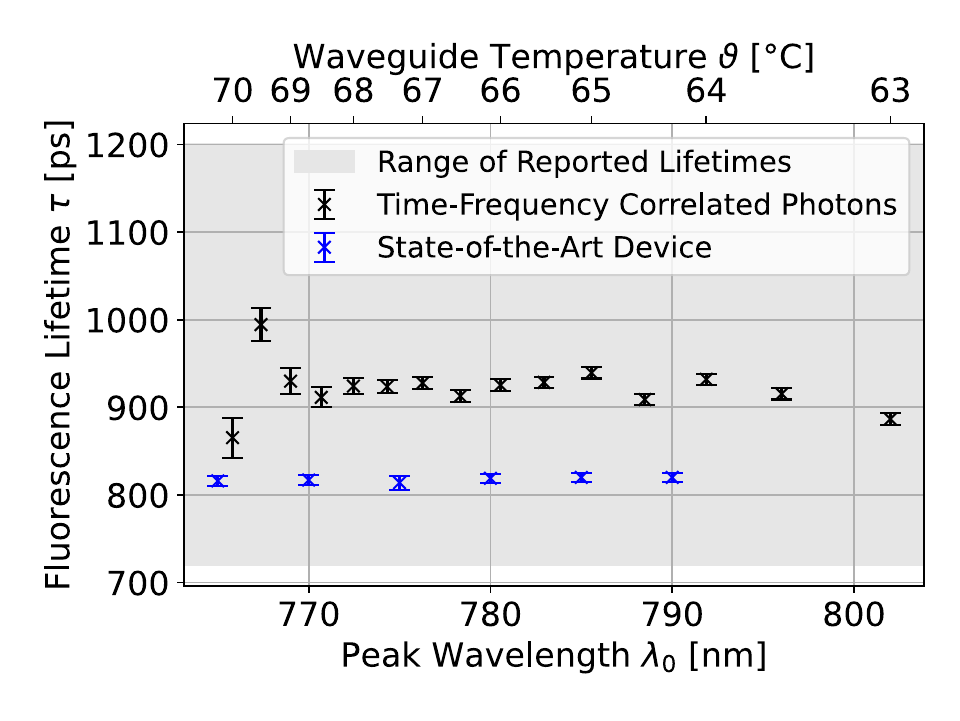}\label{fig:tau Temperature}
    \caption{Fluorescence lifetime $\tau$ depending on the waveguide temperature $\vartheta$ and the corresponding excitation center wavelength $\lambda_0$. Black indicates data gained by the described method using entangled photons, while blue are comparative values measured by a state-of-the-art FLIM. The gray background visualizes the range of reported lifetimes~\cite{Hutchinson1996,Brich1991}.}
    \label{fig:Lifetime Temperature}
\end{figure}

The fluorescence lifetime $\tau$ measured by the setup of fig.~\ref{fig:setup} depending on the waveguide temperature $\vartheta$ or rather the excitation center wavelength $\lambda_0$ are shown by the black data points in fig.~\ref{fig:Lifetime Temperature}. A clear influence of $\lambda_0$ is not visible because nearly all values of $\tau$ are on the same level. This becomes obvious because no extraordinary energy transitions are known for IR-140, which would break Kasha's rule~\cite{delValle2019}. However, strong deviations arise for temperatures above $\SI{69}{\celsius}$. Similar to the cases of short integration times $T$ or low signal-to-noise ratios $SNR$, the histograms of $IRF$ and $\widetilde{F}$ slowly become lost in noise due to the reduced photon rates, which inhibit confidence in data evaluation.

To show a significant absorption wavelength dependency of $\tau$, fluorescence samples, which allow extraordinary electronic transitions and, thus, contradict Kasha's rule, are necessary. Such samples are not available in the accessible wavelength range of our photon pair source to our knowledge. However, because of the possibility of designing photon pair sources with well-adjusted wavelengths of both correlated photons~\cite{Gilaberte2021,Torres2023}, a subsequent study could be able to show wavelength-dependent lifetimes by selecting a reasonable combination of fluorescence sample and photon pair source. 

A spectroscopic analysis of $\tau$ regarding the fluorescence emission wavelength, as it is usually done in sFLIM, is also conceivable~\cite{Fujihashi2025}. For this purpose, a monochromator can be placed in front of detection channel 2 to perform wavelength-selective measurements. However, these measurements may not yield significant coincidence histograms because of the considerable loss of fluorescence photons in the monochromator.

\subsection{Comparison with State-of-the-Art FLIM}
\label{sec:classical FLIM}

Another notable point is the difference compared to the lifetimes measured by a state-of-the-art device (blue data points in fig.~\ref{fig:Lifetime Temperature}). The underlying histogram data were measured by a Leica Stellaris 8 with a multi-color laser and a HyD R detector. The determination of $\tau$ and $\sigma_\tau$ was performed by the software Leica Application Suite X in this case instead of using eq.~\eqref{eq:IRF} and~\eqref{eq:Ftilde} since no substantial differences were expected. The data from the method using entangled photons exhibit a mean value of $\left<\tau\right>\approx\SI[separate-uncertainty = true]{932}{\pico\second}$, whereas the mean value measured by Leica Stellaris 8 is $\left<\tau\right>\approx\SI[separate-uncertainty = true]{818}{\pico\second}$. Furthermore, the values of $\tau$ shown in sec.~\ref{sec:IntTime} and ~\ref{sec:HeraldingEfficiency} also differ slightly. According to several publications, the fluorescence lifetime of IR-140 strongly varies depending on the chemical environment between approximately $\SI{200}{\femto\second}$~\cite{Wang2003} and $\SI{1.2}{\nano\second}$~\cite{Brich1991}. A reported value more comparable to our sample of IR-140 solved in ethanol is $\SI{0.72}{\nano\second}$ with excitation at $\SI{790}{\nano\metre}$ and fluorescence emission filtered by a $\SI[separate-uncertainty = true]{830(10)}{\nano\metre}$-bandpass filter~\cite{Hutchinson1996}. Consequently, these widely various lifetime values available in the literature indicate that the precise determination depends strongly on the experimental conditions, for example, on the solvent or on the environment temperature~\cite{Inada2019}, which may explain the differences in the shown data. Since we strongly assume that the visible variance in our data results from slightly changed conditions between the different measurements, this indicates that also FLIM using time-frequency correlated photons is applicable to monitor environmental influences.

However, both, the state-of-the-art and the presented setup, show results in the same order of magnitude, and within the range of previously reported values. In light of the variability in lifetime $\tau$ reported in the available literature,  this demonstrates their high comparability. This study highlights that the method using photon pairs can investigate such properties of fluorescence dyes, even though it is not yet fully technologically developed.

An important quantity in FLIM is the so-called figure-of-merit (also often called F-value) $\mathcal{F}$. It quantifies the sensitivity of a FLIM method in relation to the number of counts $N$~\cite{Gerritsen2002}.
\begin{eqnarray}
    \mathcal{F} &=& \frac{\sigma_\tau}{\tau}\cdot\sqrt{N}
    \label{eq:figure-of-merit}
\end{eqnarray}
In this definition of $\mathcal{F}$, $N$ represents the sum of all counts in a background-corrected histogram $\widetilde{F}$ over all time bins $t_i$. 
\begin{eqnarray}
    N &=& \sum_{i=0}^{N_t}\widetilde{F}\left(t_i\right)
    \label{eq:sum counts histogram}
\end{eqnarray}
In an ideal measurement, $\mathcal{F}$ is equal to unity, whereas real measurements show $\mathcal{F}>1$.

\begin{figure}[tpb]
    \centering
    \includegraphics[width=0.45\textwidth]{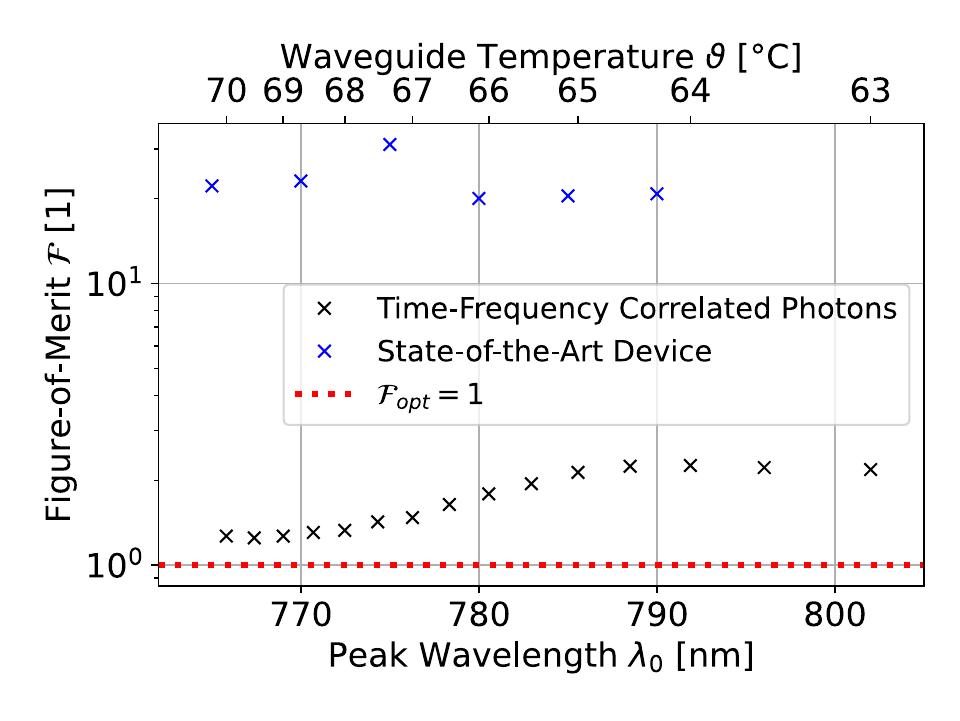}
    \caption{Figure-of-merit $\mathcal{F}$ depending on the waveguide temperature $\vartheta$ and the corresponding excitation center wavelength $\lambda_0$. Black indicates data gained by the described method using entangled photons, while blue are comparative values measured by a state-of-the-art FLIM. The red dotted line represents the optimum of $\mathcal{F}$ for an ideal measurement.}
    \label{fig:figure-of-merit}
\end{figure}

Fig.~\ref{fig:figure-of-merit} shows the figure-of-merit $\mathcal{F}$ of the data related to the measurements for the spectroscopic approach. As visible, $\mathcal{F}$ is substantially larger for the state-of-the-art FLIM. Since the values of $\tau$ and $\sigma_\tau$ in both methods are nearly identical, the different values of $\mathcal{F}$ are mainly caused by the differences regarding $N$. In classical FLIM, pulsed lasers are strongly reduced into an average power in order of $\SI{1}{mW}$ to avoid the effect of pile-up. Due to this reduction, only one event per laser pulse will be detected on average~\cite{Hirvonen2020}. However, the amount of fluorescence photons is still larger compared to FLIM with correlated photon pairs. For example, during the described experiments, the photon pair source generated less than $\SI{1}{\nano\watt}$ of SPDC power utilized for fluorescence excitation and triggering. As a result, histograms consisting of coincidences of heralding and fluorescence photons have fewer counts compared to histograms of classical time-domain FLIM, thus allowing to extract the same fluorescence lifetime information with less photons. This finally results in a substantial improvement in the figure-of-merit $\mathcal{F}$. 

In many protocols based on heralded single-photon sources measuring one photon of the pair, which results in twin photon collapse into a single-photon Fock state, ideally exhibits zero photon-number variance. Based on this, several quantum enhanced protocols have proved better-than-classical parameter estimation, for example, for absorption~\cite{Brida2010,Whittaker2017,Sabines-Chesterking2019} or phase measurements~\cite{Kuzmich1998,Pezze2008}. Accessing this quantum-enhanced regime may improve the presented method even more, but also requires low losses and high detection efficiencies~\cite{Whittaker2017}. For this reason, the implementation of such regime in FLIM with entangled photons might be challenging because of the inherent limits in single-photon fluorescence efficiency. Nevertheless, further investigations could provide more quantitative bounds to the quantum-enhanced regime.

\section{Conclusion}
\label{sec:Conclusion}

We demonstrated several experimental limits of fluorescence lifetime determination using time-frequency correlated photons. For this purpose, continuous-wave pumped source based on SPDC was used. In principle, this method can be used with a minimum number of photon pairs, which possibly reduce the risk of photobleaching in comparison to common FLIM techniques. Furthermore, considering the figure-of-merit demonstrates that measurement precision comparable to state-of-the-art devices is achievable. Only the rate of generated fluorescence photons and the integration time of the photon counting are crucial. Because of the narrow bandwidth of a waveguide-based photon pair source, our method also allows a spectroscopic approach for the analysis of molecular properties without high equipment costs and, thus, represents a conceivable alternative to state-of-the-art methods for fluorescence lifetime spectroscopy and imaging.

For application-related end-usage in biology, chemistry or medicine, we suggest several improvements. For example, a photon pair source with more narrow bandwidth will enhance the wavelength-resolution. Also the imbalance between the rate of heralding and fluorescence photons has to be improved to increase the signal-to-noise ratios and heralding efficiencies, which will enable the shortening of the required integration times. Nevertheless, the functionality of this method can be further enlarged. A scanning approach or a ghost-imaging configuration, in which the heralding detector is replaced by a camera and the image is gained by pixel-wise coincidence measurements~\cite{Scarcelli2008,Gili2022,Tian2011}, can be easily implemented to obtain images of structured samples. Moreover, an amplitude modulation of the pump laser is embeddable to allow frequency-domain fluorescence lifetime measurements. With this, FLIM devices simultaneously operable in time- and frequency-domain are conceivable, which make use of benefits of both concepts. For example, the possibility to switch rapidly between the two modes enables fast imaging using the frequency-domain mode on the one hand and the determination of multiple lifetime components using the time-domain mode on the other hand~\cite{Datta2020}.

\section*{Data Availability Statement}
\label{sec:Data Availability Statement}
All data presented in this study, as well as the underlying raw data and python codes, are openly available on Zenodo (ref.~\cite{Gaebler2025}).

\section*{Funding}
\label{sec:Funding}
This research was funded by the Bundesministerium für Bildung und Forschung (BMBF), funding programs LIVE2QMIC (FKZ: 13N15954), QC4EP (FKZ: 13N16758) and Photonics Research Germany (FKZ: 13N15713 and 13N15717). It is integrated into the Leibniz Center for Photonics in Infection Research (LPI). The LPI, initiated by Leibniz Institute of Photonic Technology IPHT, Leibniz Institute for Natural Product Research and Infection Biology (Hans Knöll Institute) HKI,  University Hospital Jena UKJ and Friedrich Schiller University Jena, is part of the BMBF national roadmap for research infrastructures. We also acknowledge the European Union's Horizon 2020 Framework program Qu-Test (grant agreement ID 101113901) for financial support.

The Leica Stellaris 8 (used for the classical fluorescence measurements in fig.~\ref{fig:LambdaLambdaScan},~\ref{fig:Lifetime Temperature} and ~\ref{fig:figure-of-merit}) is an integral part of the Microverse Imaging Center. For this, we acknowledge financial support by the Deutsche Forschungsgemeinschaft (DFG, German Research Foundation; Germany's Excellence Strategy – EXC 2051 – Project-ID 390713860; project number 316213987 – SFB 1278; instrument funding ID 460889961 multi-photon laser scanning device).

\section*{Author Contributions}

Conceptualization: T.B.G.; Methodology: T.B.G. and V.F.G.; Investigation: T.B.G., N.J. and P.T.; Software: T.B.G.; Formal analysis: T.B.G.; Data curation: T.B.G.; Writing - original draft: T.B.G.; Writing - review and editing: all authors; Visualization: T.B.G.; Supervision: M.G. and V.F.G.; Project administration: T.B.G.; Funding acquisition: M.G. and C.E.

All authors have read and agreed to the published version of the manuscript.

\section*{Acknowledgments}
\label{sec:Acknowledgments}
T.B.G. thank Nathan Harper and Scott K. Cushing (both California Institute of Technology) for fruitful discussions about the improvement of the microscope setup design and data processing.

T.B.G. thanks Kevin Lindt (TKFDM Data Steward) for support regarding topics of research data management.

We thank the Microverse Imaging Center (as well as Aurélie Jost and Sophie Neumann) for providing microscope facility support for data acquisition (and data analysis).

\section*{Conflicts of Interest}
\label{sec:Conflicts of Interest}
The authors declare no conflict of interest.

\bibliography{References.bib}

\end{document}